\pdfoutput=1
\documentclass[final,3p,times,authoryear]{elsarticle}

\usepackage{amsmath,graphicx}
\usepackage{epstopdf}
\usepackage{float}
\biboptions{sort&compress}
\usepackage{subfig}

\usepackage{amssymb,epsfig,latexsym,bm}

\usepackage{multicol}
\usepackage{color}
\usepackage{soul}

\usepackage[figuresright]{rotating}

\begin{document}

\begin{frontmatter}

\title{{\bf{\huge{A Time-Frequency Domain Approach of  Heart Rate Estimation From Photoplethysmographic (PPG) Signal}}}}
\author[buet]{Mohammad Tariqul Islam}
\ead{mhdtariqul@gmail.com,tariqul@eee.buet.ac.bd}
\author[buet,ucr]{Ishmam Zabir}
\ead{zabir173@gmail.com,izabi001@ucr.edu}
\author[buet]{Sk. Tanvir Ahamed}
\ead{tanvirahamed@outlook.com}
\author[buet]{Md. Tahmid Yasar}
\ead{ytahmid@yahoo.com}
\author[buet]{Celia Shahnaz}
\ead{celia@eee.buet.ac.bd} \cortext[corr_au]{Corresponding
    author.}
\author[buet]{Shaikh Anowarul Fattah\corref{corr_au}}
\ead{fattah@eee.buet.ac.bd }
\address[buet]{Department of Electrical and Electronic Engineering,
    Bangladesh University of Engineering and Technology (BUET), Dhaka
    1205, Bangladesh}
\address[ucr]{Department of Electrical and Computer Engineering, University of California Riverside, 900 University Ave, Riverside, CA 92521}

\begin{abstract}

\textit{Objective-} Heart rate monitoring using wrist type Photoplethysmographic (PPG) signals is getting popularity because of construction simplicity and low cost of wearable devices. The task becomes very difficult due to the presence of various motion artifacts. The objective is to develop algorithms to reduce the effect of motion artifacts and thus obtain accurate heart rate estimation. \textit{Methods-} Proposed heart rate estimation scheme utilizes both time and frequency domain analyses. Unlike conventional single stage adaptive filter, multi-stage cascaded adaptive filtering is introduced by using three channel accelerometer data to reduce the effect of motion artifacts. Both recursive least squares (RLS) and least mean squares (LMS) adaptive filters are tested. Moreover, singular spectrum analysis (SSA) is employed to obtain improved spectral peak tracking. The outputs from the filter block and SSA operation are logically combined and used for spectral domain heart rate estimation. Finally, a tracking algorithm is incorporated considering neighbouring estimates. \textit{Results-} The proposed method provides an average absolute error of 1.16 beat per minute (BPM) with a standard deviation of 1.74 BPM while tested on publicly available database consisting of recordings from 12 subjects during physical activities.  \textit{Conclusion-} It is found that the proposed method provides consistently better heart rate estimation performance in comparison to that recently reported by TROIKA, JOSS and SPECTRAP methods.
\textit{Significance-} The proposed method offers very low estimation error and a smooth heart rate tracking with simple algorithmic approach and thus feasible for implementing in wearable devices to monitor heart rate for fitness and clinical purpose.
\end{abstract}


\end{frontmatter}

\section{Introduction}

Photoplethysmography (PPG) is an electro-optic technique of measuring  various clinical observations, such as respiration rate and heart rate (\cite{bib:Allen}). Unlike electrocardiogram (ECG) signal, PPG signal can  be easily acquired by using simple pulse-oximeter and therefore, smart wearable devices use PPG to monitor heart rate. The amount of blood varies with cardiac cycle, which causes variation in the reflected light intensity in pulse-oximeter and thus the periodicity of the PPG signal corresponds to the cardiac heart rate. However, depending on the type of applications, recorded PPG data can be heavily corrupted by motion artifacts, which may affect estimation of heart rate.

Different methods have been proposed to reduce the effect of motion artifacts and determine heart rate from PPG signal. \cite{bib:Hayes} have developed a heuristic non-linear physical model to handle motion artifacts generated from probe coupling and inappropriate sensor design. \cite{bib:Patent1} have proposed a recognition scheme to identify uncorrupted pulses based on pulse size, occurrence and nature of light. The smoothed pseudo Wigner-Ville distribution is used by \cite{bib:SPWVD} to obtain heart rate estimation with increased time-frequency resolution.

Several decomposition techniques have been used to handle the effect of motion artifacts in PPG signal, such as independent component analysis (\cite{bib:ICA}), wavelet decomposition (\cite{bib:wavelet1}, \cite{bib:wavelet2}) and empirical mode decomposition (\cite{bib:EMD1}, \cite{bib:EMD2}). Different adaptive filters  are also used to reduce motion  artifacts during heart rate estimation (\cite{bib:kalman1}, \cite{bib:ANC2}, \cite{bib:NLMS}). One major concern here is to obtain proper reference signal for adaptive algorithm for artifact removal, especially when subjects are exercising. \cite{bib:NLMS} have proposed a two stage normalized least mean square (NLMS) algorithm, which can provide estimation of heart rate with high accuracy. Here, the reference signal is produced by subtracting the two channel PPG signals. \cite{bib:Spectrum} proposed spectral subtraction based motion artifact removal method, where it is shown that accelerometer data can be served as reference.

Recently, \cite{bib:TROIKA} have proposed a method namely 'TROIKA'  that utilizes signal decomposition and sparse signal reconstruction for heart rate estimation. One major step here is to make the PPG signal enough sparse so that convergence in the proposed algorithm is obtained. In this method, deviation of track of estimated heart rate from ground truth occurs when the location of spectral peaks corresponding to motion artifacts and that of the desired heart rate peak in PPG signal are in the close vicinity. In order to overcome the drawback, \cite{bib:JOSS} has introduced a method namely JOSS based on multiple measurement vector (MMV) method where a reduced sampling rate is proposed. In comparison to the results obtained by TROIKA, JOSS method offers low average error at the expense of high standard deviation. Another drawback of these methods is that they use computationally expensive M-FOCUSS algorithm to compute the spectrum. Moreover, in the JOSS method, results are reported by excluding some of the initial time windows of the dataset. The SPECTRAP (\mbox{\cite{bib:SPECTRAP}}) method, employs a post processing scheme which involves accessing the future data thereby making the method offline. In COMB (\mbox{\cite{bib:zhangcomb}}) method, the authors directly removed some noise-related IMFs, which may mix with the PPG components. Such a noise reduction scheme involves manual intervention in selecting time windows to be denoised as well as the IMFs to be discarded. Hence a method, which can provide very satisfactory heart rate estimation performance by utilizing both time and frequency domain approaches to overcome the effect of motion artifacts, is still in great demand.

In this paper, an efficient heart rate estimation scheme is proposed based on multi-stage RLS adaptive filtering and singular spectrum analysis (SSA). In the cascaded RLS filtering, three channel accelerometer data are separately used as reference to each RLS filter block. A logical combination of RLS filtered signal and SSA output are used to obtain significant reduction in motion artifact. Moreover, a tracking algorithm is performed considering previous neighbouring BPM estimations to improve the current heart rate estimation. Finally, experimentation is carried on standard PPG database and results are compared with that obtained by some recent heart rate estimation methods. Overall the contributions of the paper are,
\begin{enumerate}
\item [$\bullet$] A fast time-frequency domain denoising technique for PPG signal using multi-stage adaptive filtering in time domain and singular spectrum analysis using frequency domain has been proposed.
\item [$\bullet$] The proposed method provides very low estimation error and a smooth heart rate tracking with simple algorithmic approach.
\item [$\bullet$] The proposed algorithm is feasible for implementing in wearable devices to monitor heart rate for fitness and clinical purpose.
\end{enumerate}

\section{Proposed Method}

\begin{figure}[h]
  \centering
  \includegraphics[width=3in, height = 3.5 in]{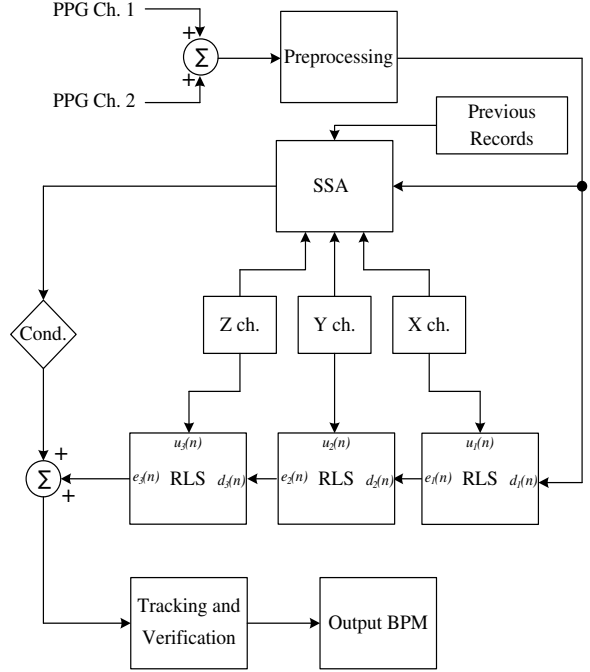}\\
  \caption{Major steps involved in the proposed method. }\label{Fig:block_diagram}
\end{figure}

The proposed method consists of three major steps: preprocessing of the given raw PPG data, noise reduction based on cascaded adaptive filtering and spectral decomposition, and spectral domain heart rate estimation. In Fig. \ref{Fig:block_diagram}, a block diagram of the proposed method is presented. At the first stage, average of the two channel PPG data (ch.1 and ch.2) is used for prefiltering. Next, in order to overcome the effect of motion artifact, two different approaches are used: one in time domain and the other one is in frequency domain. Since three channel accelerometer data are separately available, instead of employing conventional single stage recursive least squares (RLS) adaptive filter algorithm, a cascaded three stage RLS algorithm is proposed. In the figure, three channel accelerometer data are indicated by X, Y and Z channels. At the last stage, for the purpose of spectral peak estimation, the output of the adaptive filter stage is logically utilized depending on the estimate obtained from the output of the singular spectrum analysis (SSA) stage. In order to overcome the fluctuations in estimated heart rates, a weighted moving average based tracking algorithm is employed. The proposed heart rate estimation process is carried out in frame by frame basis considering overlapping frames for avoiding unwanted change in PPG and accelerometer signals.

\subsection{Preprocessing}

PPG data acquired from the two channels generally consist of underlying PPG signal, motion artifact caused by the hand swing and random noises introduced during the data recording process from the surroundings and sensor coupling (\cite{bib:TROIKA}). Because of the band limited nature of the PPG signal, bandpass filtering within the band limit $0.4$ to $3.5$ Hz is carried out on PPG signals obtained from both the channels using spectral domain filtering. In this case, the spectral coefficients beyond the specified frequency range are eliminated. For the sake of consistency, a similar bandpass filtering is performed on each of the three axis accelerometer data. 

Since the PPG sensors are placed very closely, in view of obtaining the heart rate from the acquired two PPG signals, instead of considering them separately, in this paper, an average of two channel PPG data is utilized. It is expected that due to the averaging operation, presence of unwanted random noises in different PPG channels will be reduced.

\begin{figure}[h]
  \centering
  \includegraphics[width=3.7in, height = 2.5 in]{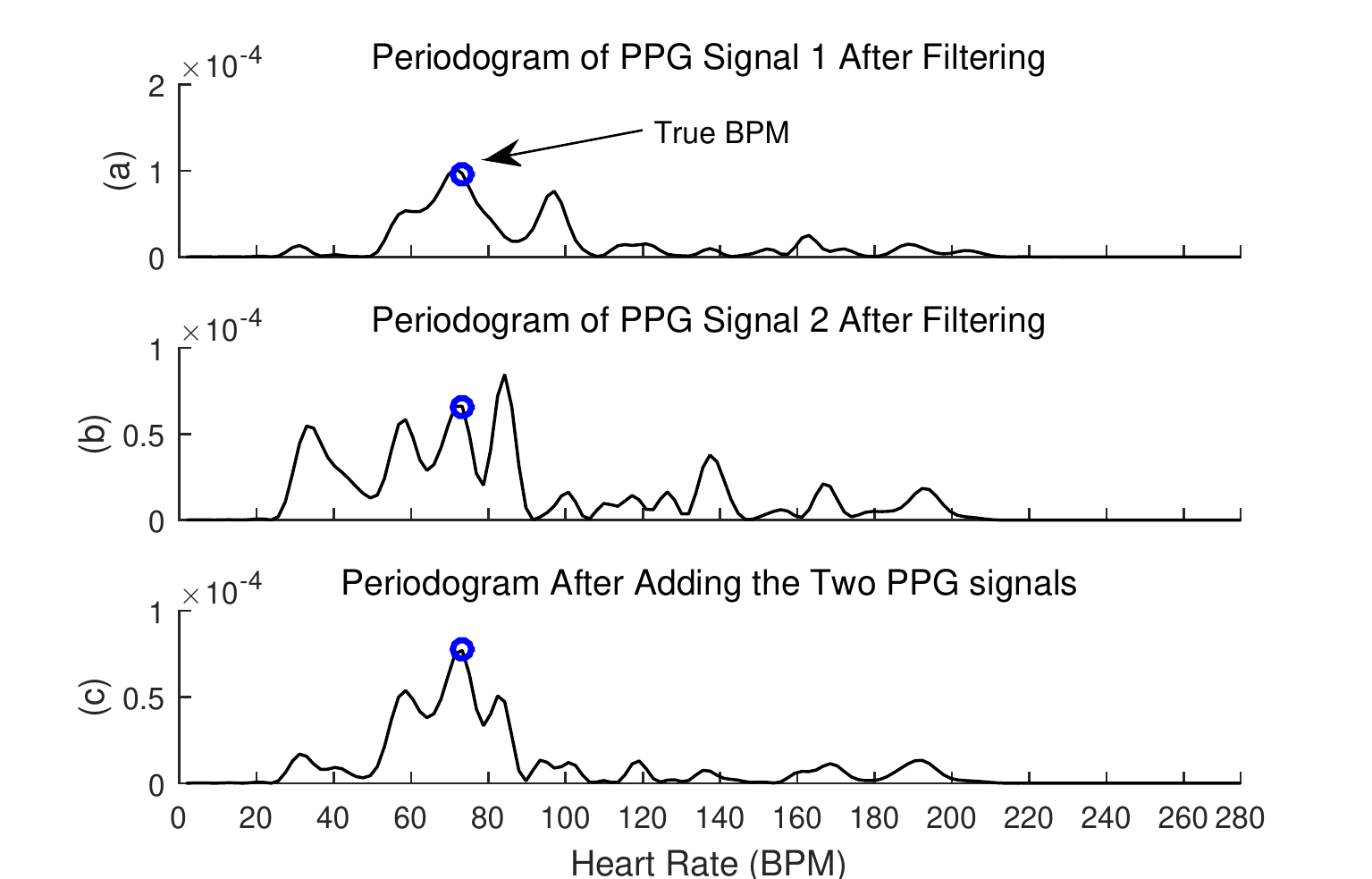}\\
  \caption{ An example showing the effect of preprocessing step (a)Periodogram of a raw PPG signal of channel-1 (frequency corresponds to heart rate is unique)  (b) Periodogram of raw PPG signal of channel-2 of the same time window (frequency corresponds to heart rate is not distinct)  (c) The two PPG signals of two channels of the same time window are pre-filtered and averaged in time domain. Periodogram of the obtained signal exibit a distinct peak at true heart rate.}\label{Fig:prefiltering}
\end{figure}

In Fig. \ref{Fig:prefiltering}, the effect of proposed prefiltering and channel-averaging stages are presented using sample PPG data in spectral domain. In Fig. 2(a) and 2(b), periodograms of PPG data obtained from two channels (at the same time instant) are shown. In these figures, X-axis (frequency axis) corresponds to heart rate in BPM. For the purpose of clear visibility, data from each channel are band pass filtered (0.4-3.5 Hz). It is clearly observed that PPG signal from channel-1 exhibits clearly distinguishable peak whereas in the periodogram of channel-2 data, closely spaced competing peaks exist. True BPM is marked in each periodogram. Use of channel-2 data may provide misleading heart rate detection. This may happen to channel-1 data for some other time instants. Hence, use of only one channel out of two channels is always very risky. In the proposed method, average of two channel data is utilized. In Fig. 2(c), effect of averaging the two channel data on the resultant periodogram is shown. Because of averaging, significant noise reduction is achieved with prominent PPG peak. After the preprocessing stage, the presence of noise like characteristics in the PPG signal is mainly due to the motion artifacts. In what follows, a method is proposed to reduce the effect of motion artifacts.

\subsection{Multi-stage Cascaded RLS Filtering}

\begin{figure}[h]
	\centering
	\includegraphics[width=3in, height = 3in]{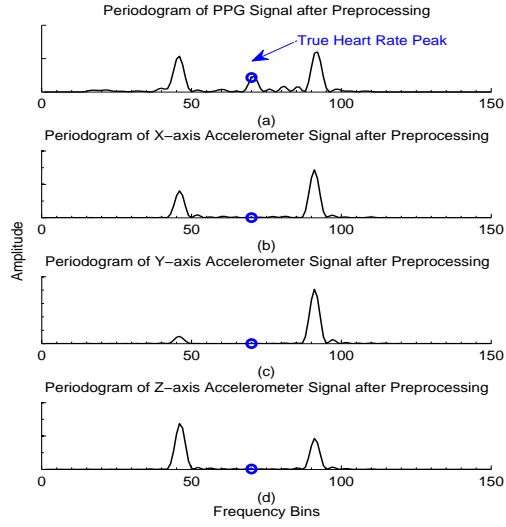}\\
  \caption{(a) Periodogram of PPG signal after filtering and adding (though frequency corrresponds to heart rate is separable, it is not distinct). (b), (c) and (d) represent peridogram of X, Y and Z-axis accelerometer signal respectively}\label{Fig:signal}
\end{figure}
 
Presence of motion artifacts in the acquired PPG data plays a detrimental role in PPG based heart rate detection, especially when significant physical movement is involved during the PPG data recording. In cases where motion artifacts are present predominantly, their characteristics can be convincingly captured from the accelerometer data. In order to demonstrate the mapping between the three channel accelerometer data and the motion artifacts present in the acquired PPG data, in Fig. \ref{Fig:signal} (a), spectral representation of the pre-processed PPG data obtained in the previous sub-section is shown. In this figure, the spectral value corresponding to the ground truth heart rate obtained from the ECG data is marked by a blue circle. Here apart from the peak corresponding to the true heart rate (referred to desired peak), some other peaks caused by the motion artifacts (referred auxiliary peaks) are also clearly visible. In Figs. \ref{Fig:signal}(b), (c) and (d), periodograms corresponding to the three channels of accelerometer are shown. It is observed that the peaks present in Figs. \ref{Fig:signal}(b), (c) and (d) can easily be related to the auxiliary peaks appeared in Fig. \ref{Fig:signal}(a). In different time instances, the three different channels of accelerometer capture different effects caused by the motion artifacts which may be visible in the pre-processed PPG data. Hence, in order to extract complete information regarding the motion artifacts, unlike some methods available in literature, it would not be sufficient to consider any one of the three channel data or just the average of the three channel data. In its simplest form, one may represent the motion artifacts present in the preprocessed PPG signal in terms of linear combination of the three components as
\begin{equation}
\boldsymbol{m}[n] = a_x\boldsymbol{m_x}[n] + a_y\boldsymbol{m_y}[n] + a_z\boldsymbol{m_z}[n],
\end{equation}

\begin{figure}[h]
  \centering
  \includegraphics[width=3in, height = 1.3 in]{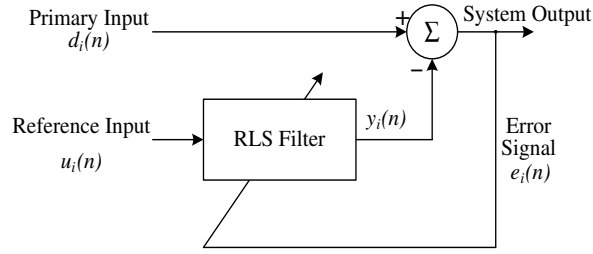}\\
  \caption{Block diagram of single (i-th) stage of adaptive RLS noise canceller. }\label{Fig:afbd}
\end{figure}

\begin{figure}[h]
	\centering
	\includegraphics[width=3in, height = 4in]{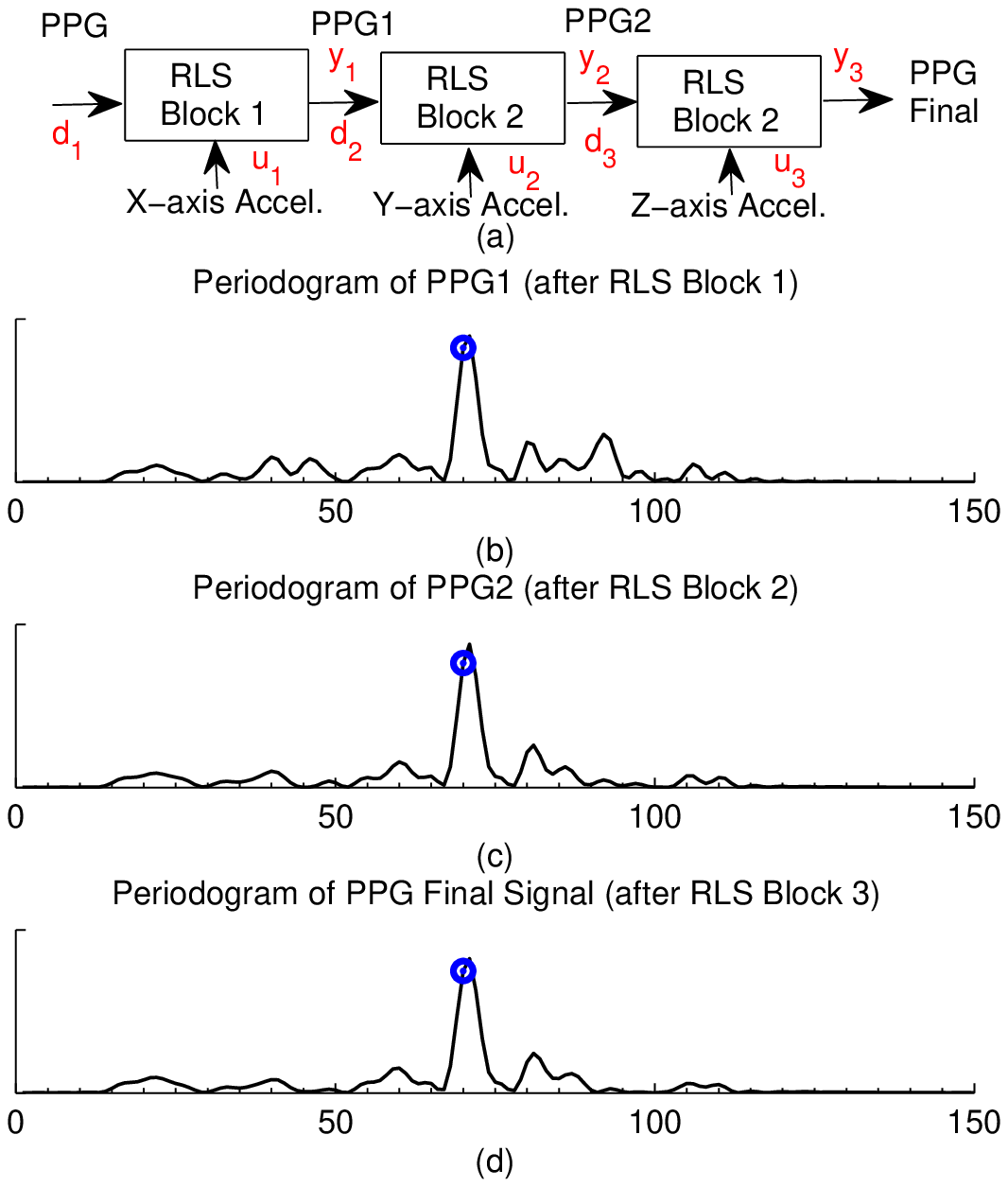}\\
  \caption{An example of step by step motion artifact cancellation, (a) Block diagram of proposed three stage RLS filter. (b), (c) and (d) represent peridogram of PPG signal after each RLS block. The step by step cancelletion of  motion artifacts by RLS filtering gives a distinct peak at true heart rate frequency.}\label{Fig:stepbystep}
\end{figure}

where $\boldsymbol{m_x}[n]$, $\boldsymbol{m_y}[n]$ and $\boldsymbol{m_z}[n]$ are assumed to be motion artifacts due to the movements along X, Y and Z-axis and $a_x$, $a_y$ and $a_z$ are corresponding constants. The data acquired from the three different channels of accelerometer can be approximated as equivalent to components of motion artifacts. Hence, in the proposed method, accelerometer data from all three channels are separately taken into consideration. For the purpose of removing motion artifacts from the preprocessed PPG data, a multi-stage adaptive filtering based scheme is proposed where the adaptive filtering operation is carried out in three separate stages considering three channel accelerometer data separately as reference in those three stages. Because of the robustness and faster implementation, the recursive least squares (RLS) adaptive filter algorithm is chosen in the proposed method. In Fig. \ref{Fig:afbd}, block diagram of a single stage RLS adaptive filter is shown, where $\boldsymbol{d_i}(n)$ represents the input signal (at the i-th stage) containing the desired signal corrupted by noise and $\boldsymbol{u_i}(n)$ represents the noise reference (at the $i$-th stage), which is one of the three accelerometer data. In \ref{Fig:stepbystep}(a), a schematic diagram of the proposed cascaded three stage RLS adaptive noise canceller is shown. Here in the first stage, $\boldsymbol{d_1}$ corresponds to preprocessed PPG signal ( with motion artifacts). As reference to the each stage of RLS adaptive filtering, X,Y and Z three channel accelerometer data are independently used as reference $\boldsymbol{u_1}$,$\boldsymbol{u_2}$ and $\boldsymbol{u_3}$, respectively in stage-1, stage-2 and stage-3. The RLS filtered output $\boldsymbol{y_1}$ at stage-1 is  used as input $\boldsymbol{d_2}$ to the next stage and the output $\boldsymbol{y_2}$ at stage-2 is used as input $\boldsymbol{d_3}$ to the stage-3. In the first stage of RLS filtering operation, it is expected that the effect of motion artifacts correspond to X-channel will be partially removed. In the second stage of RLS filtering, because of use of Y-channel accelerometer data as the reference signal, corresponding motion artifact is expected to be removed. In a similar way, in the third stage of RLS filtering, complete removal of the motion artifact is expected. In Figs. \ref{Fig:stepbystep} (b), (c) and (d), periodograms obtained at each stage are shown.

Given the desired signal $\boldsymbol{d_i}(n)$, reference signal $\boldsymbol{u_i}(n)$ and current estimated value of the filter coefficient $\boldsymbol{\hat{w}_i}(n)$, the working formulas for obtaining the updated estimate of the filter co-efficient can be expressed as (\cite{bib:book1}):


\begin{equation}
\hat{\boldsymbol{w}}_i(n) = \hat{\boldsymbol{w}}_i(n-1) + \boldsymbol{k}_i(n) e_i(n) \label{EQ:rls_w_eq}
\end{equation}
\begin{equation}
e_i(n) = d_i(n) - \hat{\boldsymbol{w}}_i^T(n-1) \boldsymbol{u}_i(n) \label{EQ:rls_err}
\end{equation}
\begin{equation}
\boldsymbol{k}_i(n) =  \frac{\boldsymbol{\pi}_i(n)}{\lambda + \boldsymbol{u}_i^T(n) \boldsymbol{\pi}_i(n)} \label{EQ:rls_gain}
\end{equation}
\begin{equation}
\boldsymbol{\pi}_i(n) = \boldsymbol{P}_i(n-1)\boldsymbol{u}_i(n) \label{EQ:rls_iv}
\end{equation}
\begin{equation}
\boldsymbol{P}_i(n) = \lambda ^ {-1} \boldsymbol{P}_i(n-1)- \lambda ^ {-1} \boldsymbol{k}_i(n) \hat{\boldsymbol{u}}^T(n) \boldsymbol{P}_i(n-1)   \label{EQ:rls_iam}
\end{equation}
Here, the inverse of the covariance matrix $\boldsymbol{P}_i(n)$ and weight vector $\boldsymbol{\hat{w}}_i(n)$ need to be initialized along with a chosen value of filter length $M_{RLS}$ and forgetting factor $\lambda$.
The update equation for the weight vector is shown in Eq. {\ref{EQ:rls_w_eq}} where the weight vector is incremented from its old value by an amount equal to the multiplication of error $e_i(n)$ and gain vector $\mathbf{k}_i(n)$, which are calculated from Eqs. {\ref{EQ:rls_err}} and {\ref{EQ:rls_gain}}, respectively. The intermediate vector $\boldsymbol{\pi}_i(n)$, evaluated in Eq. {\ref{EQ:rls_iv}}, is used to calculate the gain vector. Equation {\ref{EQ:rls_iam}} sets up $\mathbf{P}_i(n)$ for the next iteration.

\begin{figure}[h]
  \centering
  \includegraphics[width=3.5in, height = 2 in]{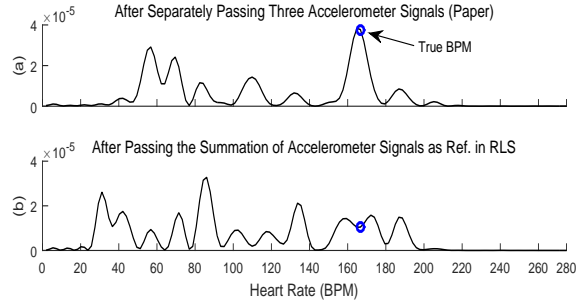}\\
  \caption{ Comparison between proposed multi-stage RLS filtering and single stage RLS filtering, (a) Peridogram of PPG signal obtained after using proposed multi-stage RLS noise canceller (true and estimated heart rate coincide), (b) Periodogram of PPG signal obtained after using single stage RLS filter with sum of 3 channel accelerometer data (false peak estimated).}\label{Fig:sumworse}
\end{figure}

In order to demonstrate the superiority of the filtered signal obtained by using the proposed method of multi-stage RLS filtering in comparison to that obtained by using the single stage filtering, in Fig. \ref{Fig:sumworse}, corresponding periodograms of filtered signals are shown. In case of single stage filtering, average of the three channel accelerometer data is used as reference signal. In Fig. \ref{Fig:sumworse}(a), it is observed that in the location corresponding to true heart rate value, there exists clearly distinguishable peak. However, in Fig. \ref{Fig:sumworse}(b) several competing peaks appear in vicinity of the desired location which may lead to poor estimation of true heart rate. 

Although in most of the cases, in the proposed multi-stage RLS filtered output, the effect of motion artifacts is significantly reduced, there are some cases where complete removal of motion artifacts may not be achieved. One possible reason behind the fact is that, the RLS filtering operation is carried out in temporal domain, where there is no scope of utilizing the spectral characteristics. It would be interesting if the spectral behavior of the available reference data from the accelerometer could be utilized in motion artifacts removal. In what follows, a spectral domain noise removing algorithm is presented, which is incorporated with a view to achieve further reduction in motion artifacts.

\subsection{SSA Based Noise Reduction}

It is found that the proposed multi-stage RLS filtering operation exhibits satisfactory performance when the motion artifacts present in the recorded PPG signal closely resemble at least one of the three accelerometer signals. This expected resemblance is found true in most of the time frames. However, due to the presence of rapid and/or huge hand swing during extensive exercise, the accelerometer signals may exhibit significant difference with the motion artifact present in the recorded PPG signal. In this case, considering accelerometer signals as the reference signals in the multi-stage RLS filtering, effective removal of motion artifact may not be possible. As an alternative, one may  decompose the PPG signal and identify motion artifact by comparing the spectral behavior of the decomposed signals with that of the accelerometer signals. 

For the purpose of comparison, instead of using the entire PPG signal, if a component signal obtained via decomposition is used, it is expected that comparatively better spectral matching can be obtained. For example, when a decomposed component corresponds to motion artifact, its spectral characteristics will show close resemblance with that of the accelerometer signals and in this way, the components corresponding to the motion artifacts can be identified and removed from the PPG signal. Among different signal decomposition techniques, in the proposed method, apart from multi-stage RLS filtering, singular spectrum analysis based motion artifact removal scheme is employed. The SSA is a well known decomposition method, which utilizes singular value decomposition (SVD) to extract a number of components (\cite{bib:book2}).

In order to investigate the spectral characteristics of the preprocessed PPG signal, instead of considering the entire signal, first the SVD is employed to decompose it into $d$ number of components. Next, a clustering scheme is used to obtain $g$ ($g \leq d $) number of groups where each group contains components exhibiting similar characteristics.  For a given signal of length $N$, in order to carry out the SVD operation, an L-trajectory matrix of dimension  $L\times K$ ($K = N - L + 1,L < N/2$) is formed. After SVD, clustering of singular values is performed based on the principle described in (\cite{bib:book2}, pp. 66) to obtain $g$ number of time series.
Next, dominant spectral peaks in each time series are determined using the periodogram, where peaks with amplitude larger than $50\%$ of the maximum amplitude are treated as dominant peaks. The idea here is to find out the time series containing dominant frequencies which are common to the dominant spectral peaks present in the accelerometer signals. For this purpose, the dominant spectral peaks in each of the three accelerometer signals are determined by using the periodogram. It is expected that the time series, whose dominant frequencies are similar to spectral peaks appearing in the accelerometer signals, appear in the original PPG signal due to the presence of motion artifacts and noise. Hence these types of time series are discarded and by using the remaining time series a comparatively enhanced PPG signal is generated. It is to be noted that, some selected dominant frequencies may be close to the fundamental and harmonic frequencies of heartbeat. As an estimate of heartbeat is available from the previous time window, the signal components containing dominant spectral peaks associated with the neighborhood of that estimated heartbeat are not discarded in the reconstruction stage. The extent of neighborhood is chosen as a very close vicinity ($\pm \delta$) of the estimated heartbeat frequency value of previous window.

\subsection{Logical Combination of Time and Frequency Domain Approaches}

\begin{figure*}[ht]
\begin{subfloat}
  \centering
  \includegraphics[width=2.2in, height = 3in]{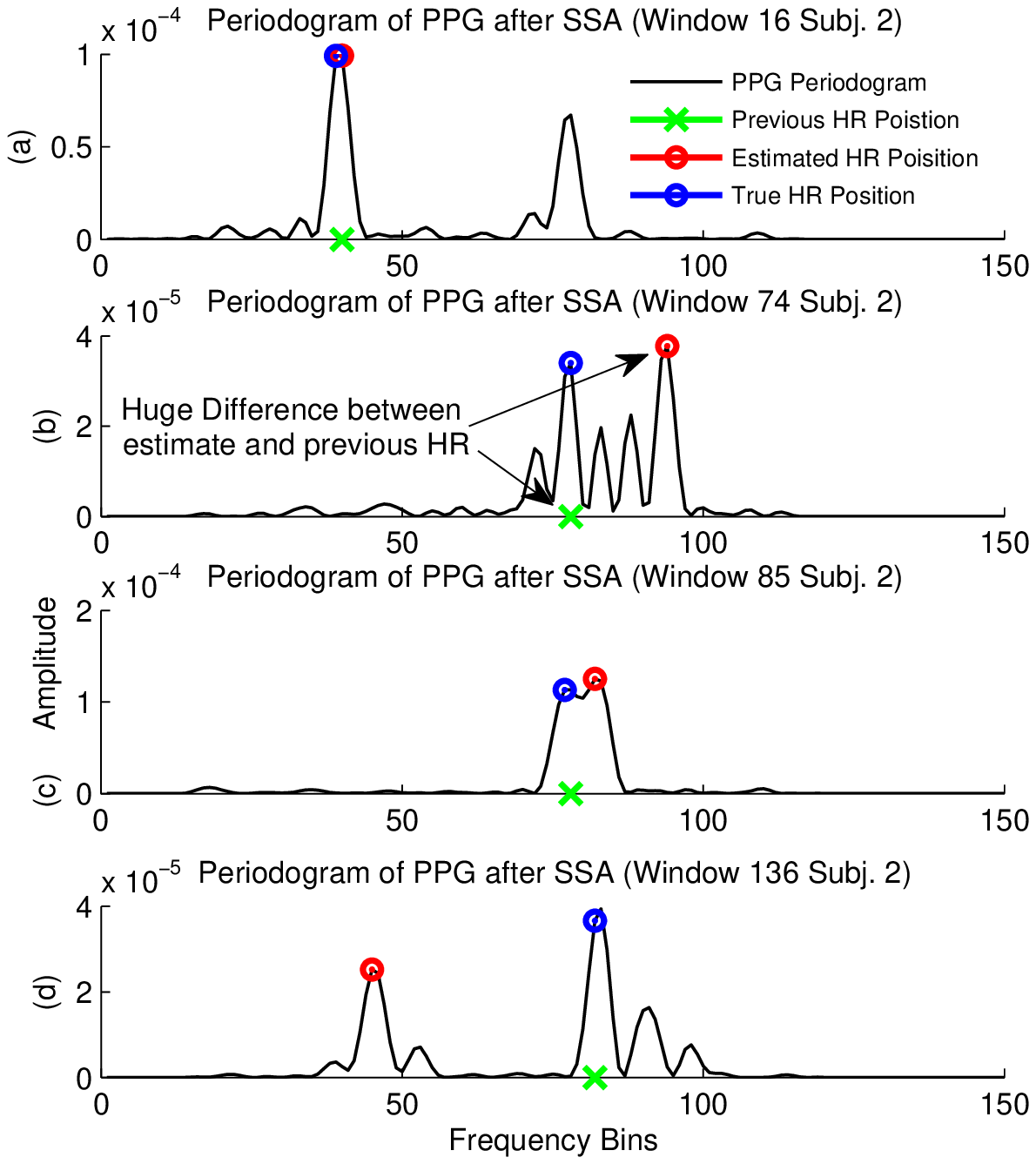}
\end{subfloat}
\begin{subfloat}
  \centering
  \includegraphics[width=2.2in, height = 3in]{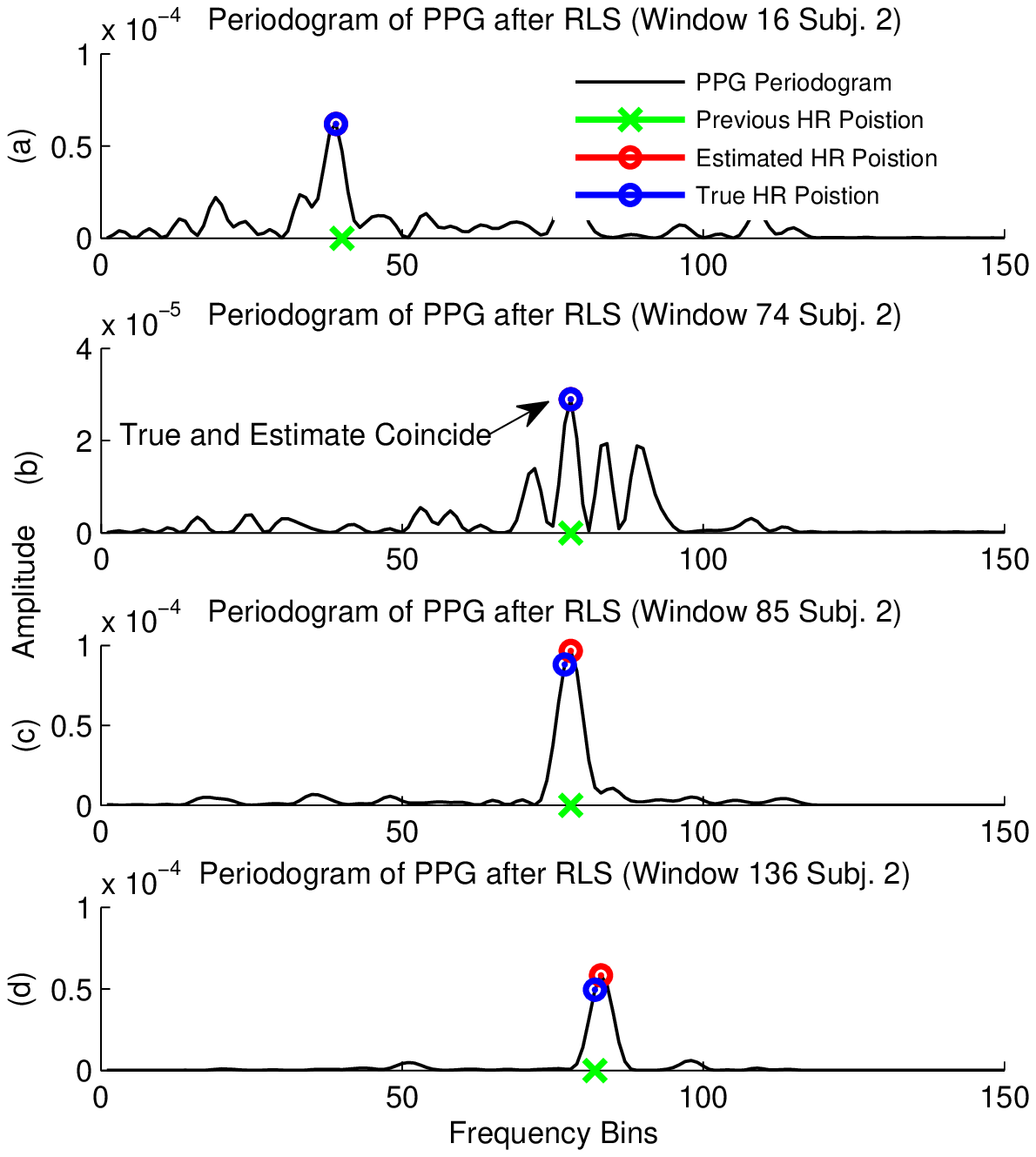}
\end{subfloat}
\begin{subfloat}
  \centering
  \includegraphics[width=2.2in, height = 3in]{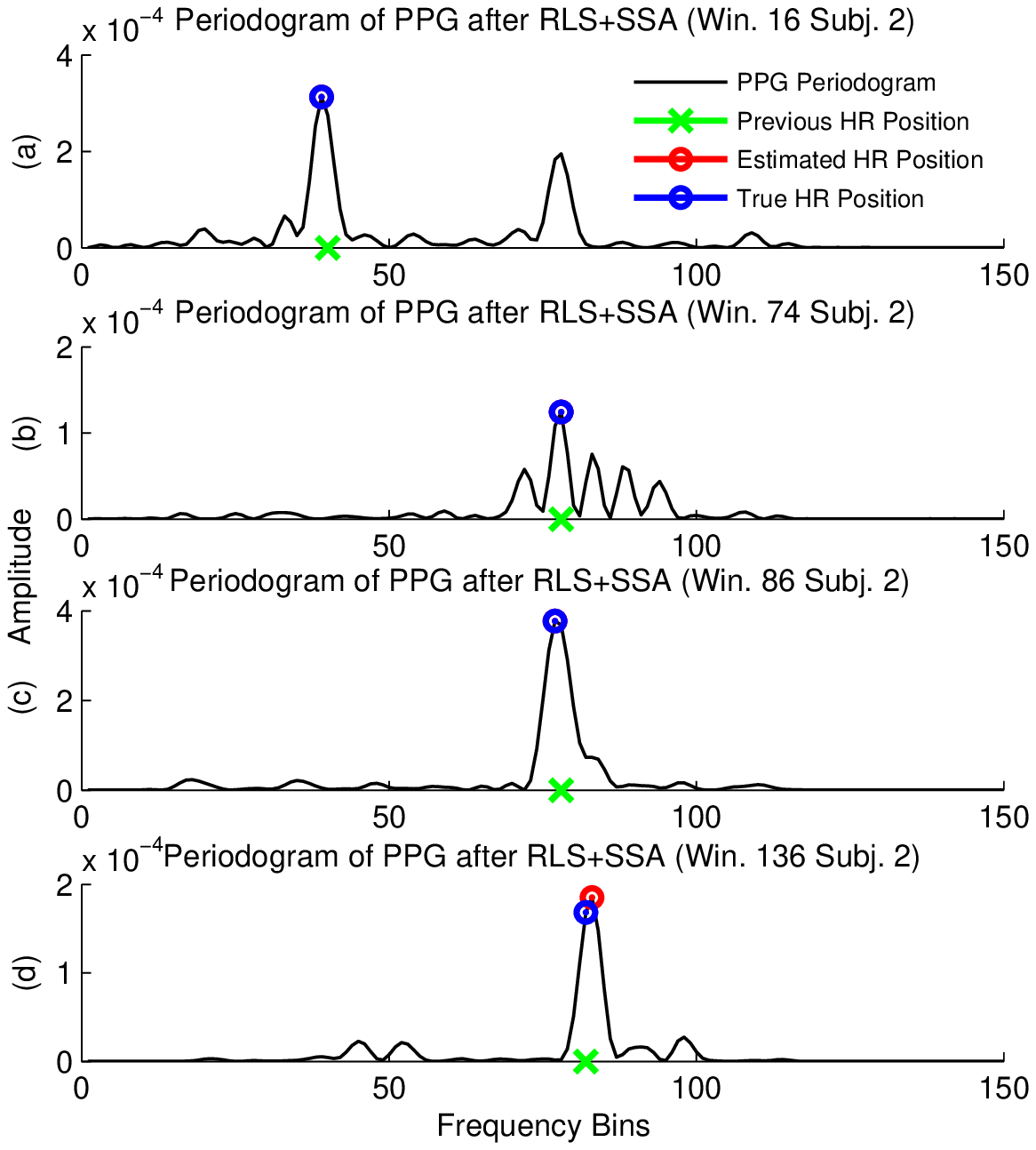}
  \caption{ Heart rate estimation using only SSA step (leftmost column), only RLS step (middle column) and proposed method (rightmost column) are shown respectively from leftmost to rightmost figures. In each of these three cases, four time windows from subject 2 are considered, (a) 16, (b) 74, (c) 85 and (d) 136. These figures in topmost row (or first row) correspond to window 16 (all labelled as 'a'). In (a), successful estimation of true heart rate by SSA and proposed method, (b) noise frequency is dominant over true heart rate frequency, (c) inseparable heart rate and noise frequency, (d) incorrectly measured heart rate due to failure in tracking of heart rate from SSA processed signal. 
 }
  \label{Fig:RLS_SSA}
  \end{subfloat}
\end{figure*}

From the preprocessed PPG signal, in order to reduce the effect of motion artifacts and noise, two different schemes are employed, the first one is the proposed multi-stage RLS filtering and the next one is SSA based noise cancellation. It is expected that the  
multi-stage RLS filtering can provide significant reduction in motion artifacts, especially when the characteristics of motion artifacts in PPG signal exhibit similarity with that of the accelerometer signals. It refers to the case where spectral peak corresponding to true heart rate is distinctly separable from dominant spectral peaks caused by motion artifacts, which also appear in spectrum of accelerometer signals. This scenario is observed in most of the cases and an example case is demonstrated in Fig. \ref{Fig:RLS_SSA}. However, RLS based noise subtraction may cause decrease in spectral energy at the true heart rate frequency in cases where the spectrum of accelerometer signals contains significant energy around the true heart rate frequency. 
In parallel to multistage RLS filtering, SSA based noise reduction scheme is implemented to suppress the energy caused by motion artifacts at the frequencies which are not in the neighbourhood of the frequency of the previous estimated heart rate. Hence, it is expected that the two noise reduced signals, one that is obtained after multistage RLS filtering and the other one obtained after SSA based noise reduction, if added together, the resulting signal may provide significant enhancement of the spectral peak corresponding to true heart rate.

However, there exists some cases where the peaks corresponding to motion artifacts have much higher amplitude than the peaks corresponding to the true heart rate in the preprocessed PPG spectrum. In this case, the suppression of energy due to motion artifacts by SSA based noise reduction is not sufficient enough and the peaks due to motion artifacts will be still higher than true heart rate peak in the spectrum of SSA processed PPG signal. Thus adding the SSA processed PPG signal ($x_S[n]$) with multistage RLS processed PPG signal ($x_{R}[n]$) may not provide expected enhancement of spectral peak at true heart rate frequency. It may also lead to false estimation of the true heart rate due to dominant nature of motion artifact in the spectrum of the two added signals. Thus a conditional sum is proposed to deal with this situation. In each time window, after obtaining $x_S[n]$, location of maximum peak is determined from the spectrum of that signal and corresponding BPM ($B_S$) is computed. If $B_S$ is within a certain range of the estimated BPM obtained in the previous time window ($B_{Prev}$), it is expected that $x_S[n]$ is motion artifact suppressed and preserves the peak due to true heart rate. Hence, adding $x_S[n]$ with $x_R[n]$ can provide desired peak enhancement. For the cases where $B_S$ is not within the certain range ($\epsilon$) of $B_{Prev}$, only $x_R[n]$ is considered to be the final PPG signal ($y_f[n]$). Such a conditional operation can be expressed as 

\begin{equation}
y_{f}[n]= 
\begin{cases}

\hat{x}_{R}[n],\ \text{if} \ B_{S}-B_{prev} \geq  \epsilon
\\
\hat{x}_{R}[n] + \hat{x}_{S}[n],\ \ \ \ \ \ \text{otherwise}
\end{cases}
\end{equation}
where, $\hat{x}_R[n]$ and $\hat{x}_S[n])$ are energy normalized versions of $x_R[n]$ and $x_S[n]$, respectively. It is to be mentioned that conditional sum operation is not performed for the first two time windows. Both multistage RLS filtering and SSA based noise reduction schemes are independently tested on several time windows containing various types of motion artifacts and also the effect of proposed conditional sum scheme is investigated. It is found that the conditional sum can provide significant  enhancement of peak corresponding to true heart rate. In order to demonstrate the performance of the proposed conditional sum scheme, four different time windows ( 16, 74, 85 and 136) from a specific person (subject-2) are taken into consideration and corresponding spectral representations are presented in Fig. \ref{Fig:RLS_SSA} (a), (b), (c) and (d) . In Fig. \ref{Fig:RLS_SSA}, from leftmost to rightmost figure, three periodograms corresponding to $\hat{x}_{S}[n]$, $\hat{x}_{R}[n]$ and  $y_{f}[n]$ are shown, where x-axis is scaled to represent heart rate in BPM. In each case, the peaks corresponding to estimated BPM and true BPM are marked using blue and  red circles respectively. BPM obtained in previous time window is also marked by green cross mark on x-axis.  It is observed from the Fig. \ref{Fig:RLS_SSA} (a) that, $B_S$ is very close to $B_{Prev}$ and thus the periodogram corresponding to the combined signal exhibits significant enhancement of the true peak. However in Fig. \ref{Fig:RLS_SSA} (b) and (d), $B_S$ is far from $B_{Prev}$ and thus as per the proposed conditional sum criterion, only $x_{R}$ is considered as $y_f[n]$.

\subsection{Heart Rate Tracking}

Tracking the heart rate is one of key steps in the proposed method. Tracking consists of three main parts: initial estimation, estimation of current heart rate and verification. Based on experimentation on several time windows, it is observed that due to the presence of motion artifacts in some cases error in estimated BPM may occur. Two major reasons behind such errors are the presence of strong peaks caused by motion artifact relatively at far location from the true heart rate peak and presence of peaks caused by motion artifacts in the close vicinity of true heart rate peak. It is well known that the general pattern of heart rate seldom exhibits random fluctuation, it either gradually increases or gradually decreases depending on human action. Considering this fact and to reduce the computational burden, instead of allowing multiple candidates (dominant peaks), in a given time window only single peak is chosen as the desired peak. For selecting such a single peak location, frequency location of the estimated heart rate of the previous time window ($N_0$) is used as a reference and a search region $P_0  =[ N_0-\Delta_s, . . ., N_0+\Delta_s ]$, that is $\pm\Delta_s$ frequency bins within the neighborhood of the reference $N_0$ is utilized. The highest peak location $N_{cur}$ is selected from that search range ($N_{cur} \in P_0$) and then the estimated BPM is calculated as 
\begin{equation}
 \hat{B}_{est} = \frac{N_{cur}-1}{N_{F}}  \times 60  \times F_s,
\end{equation}  
where $N_{F}$ is number of points used for computing periodogram and $F_s$ is the sampling frequency. 

 Next in order to obtain a smooth heart rate tracking, a three point moving average is employed where the estimate of BPM in current window is computed as  
\begin{equation}
B_{est}' = \alpha \hat{B}_{est} + \beta B_{-1} + \gamma B_{-2},
\end{equation}
where $B_{-1}$ and $B_{-2}$ represent estimated BPM of previous two successive time windows and $\alpha+ \beta + \gamma = 1$ and values of weighting parameters are empirically chosen. Finally  in order to prevent extremely high or low estimated values of BPM in comparison to previous BPM estimate, the final BPM value is computed as 
\begin{equation}
B_{est}= 
\begin{cases}
B_{-1}+\lambda_{inc},\ \text{if} \ B_{est}'-B_{-1} \geq  \lambda_{inc}
\\
B_{-1}-\lambda_{dec},\ \text{if} \ B_{est}'-B_{-1} \leq  \lambda_{dec}
\\
B_{est}',\ \ \ \ \ \ \text{otherwise},
\end{cases}
\end{equation} 
where the justifying constants $\lambda_{inc}$ and $\lambda_{dec}$ are set empirically to small integer values.

\section{Results and Performance Measurement}

\subsection{Data Acquisition}
Performance of the proposed method is tested on publicly available PPG dataset (\cite{bib:TROIKA}). In the dataset, two channel PPG signals and accelerometer signals of three axes (X,Y and Z) are available along with the ground truth values of heart rate in BPM. The data set contains signals collected from 12 male subjects where the PPG and accelerometer signals were recorded from the wrist using 125 Hz sampling rate. The duration of recording of each subject is 5 minutes where the speed of walking or running on a treadmill varies with variable time duration.

\subsection{Parameter Measurement}

In the design of RLS filter, filter length $M_{RLS}$ is chosen as $55$, the forgetting factor $\lambda$ is selected as 0.999 and weight vector $\hat{\boldsymbol{w}}_i(n)$ is initialized to zero. The inverse of the covariance matrix $P_i(n)$ is initialized as $P_i(n)= 10I$, where $I$ is identity matrix of order $M_{RLS}$.
In the SSA based noise reduction scheme, the value of time window L is chosen
400 as suggested by \cite{bib:TROIKA}. Frame by frame analysis is carried out with a frame duration of $8$ sec. which provides $1000$ samples. The search domain $\Delta_s$ for the neighbourhood of estimated heart rate of previous time window is reasonably set to $10$. In the conditional stage, the value of controlling parameter $\epsilon$ is chosen as 15, as in general, a significant change of BPM in successive frames does not occur. Considering large value of $\epsilon$ may cause severe false peak estimation which can affect the overall tracking performance. In order to obtain smooth heart rate tracking, the moving average parameters are chosen in such a way that more emphasis is given on the current estimate and relatively less on the estimates estimates of the previous two time windows. In order to obtain this, the parameters $\alpha$, $\beta$ and $\gamma$ are set to $0.90$, $0.05$ and $0.05$, respectively. Moreover, it is observed from ground truth heart rate data is that the increase of heart rate during exercise is relatively faster than the decrease of heart during resting period. So, keeping this in mind to compute the final value of beat rate using (10), the constants are chosen as $\lambda_{inc} = 5$ and $\lambda_{dec} = 3$.

In order to evaluate the performance of the proposed method, different performance indices are taken into consideration, such as average absolute error (AAE), pearson correlation ($r$) and Bland-Altman plot (\cite{bib:Bland}). The AAE is defined as 
\begin{equation}
AAE = \frac{1}{N_f} \displaystyle\sum\limits_{i=1}^{N_f} | B_{est}(i)-B_{true}(i) |,
\end{equation}
where $B_{est}(i)$ and $B_{true}(i)$ are estimated and true values of heart rate in BPM of the $i$-th frame, respectively  and $N_f$ is the total number of time frame. Pearson correlation is a measure of degree of similarity between true and estimated values of heart rate. Higher the value of $r$ better the estimates. The Bland-Altman plot measures the agreement between true and estimated values of heart rate. Here limit of agreement (LOA) is computed using the average difference ($\mu$) and the standard deviation ($\sigma$), which is defined as $[\mu - 1.96\sigma, \mu + 1.96\sigma]$.

\subsection{Results}

Different state of the art heart rate estimation methods have been implemented and results are compared with that obtained by the proposed method. Comparison methods reported by \cite{bib:TROIKA}, \cite{bib:JOSS}, \cite{bib:SPECTRAP}, and \cite{bib:zhangcomb} are referred to as TROIKA, JOSS, SPECTRAP and COMB, respectively. It is to be mentioned that three major contributions of the proposed method are: use of cascaded RLS adaptive filter, logical combination of RLS and SSA processed signals and efficient heart rate tracking. Hence, it would be interesting to compare the performance of the proposed method considering following cases:\\
C-1: SSA+ proposed tracking algorithm\\
C-2: RLS+ proposed tracking algorithm\\
C-3: Logical combination of proposed cascaded LMS+ SSA+ tracking (alternative proposed method)\\
C-4: Logical combination of proposed cascaded RLS+ SSA+ tracking (proposed method)\\
Obviously one most logical alternative to the RLS filter would be the LMS filter. In case C-3, the proposed cascaded three stage adaptive filtering with separate accelerometer reference will be used but replacing RLS filters by LMS filters. However, to the best of our knowledge, the case C-3 is also a new approach and can be considered as an alternative proposed method (proposed-2).

\setlength{\arrayrulewidth}{.1mm}
\setlength{\tabcolsep}{6.5pt}
\renewcommand{\arraystretch}{1.3}

\begin{table*}[t]
  \centering
\caption{Performance comparison among different approaches in terms of AAE considering all 12 subjects.}
  \label{table:comparison}
  \begin{tabular}{c|c|c|c|c|c|c|c|c|c|c|c|c|c}
  
    \hline
     Subject No. & 1 & 2 & 3 & 4 & 5 & 6 & 7 &  8 & 9 & 10 & 11 & 12& Average\\
    
   \hline
    
    C-1(SSA+track.) & 2.81  & \textcolor{red}{43.57} &  \textcolor{red}{50.08} & 1.52 & 0.77 & \textcolor{red}{54.68} & 1.65 & 1.50 & 0.73 & 6.19 & 5.22 & 2.79 & \textcolor{red}{14.29}\\
    
    \hline
    C-2(RLS+track.)   &     2.38  &	3.87 &	1.16 &	1.11 &	1.61 &	2.13 &	2.39 &	 0.98	& 1.17 & 	6.41 &	3.33 &	1.41 & 2.33\\

    \hline
    
    C-3(LMS+SSA+track.) & 1.82 &   1.65 &  0.84 &  1.23 & 1.23 &  1.58 &  0.99 &	0.76 &	0.64 &	4.95 &	3.03 &   2.22 & 1.75\\
    
    \hline
    
        C-4(RLS+SSA+track.) &     1.16  &  1.16  &  0.79  &  0.87    & 0.71  &  1.14  &  0.71  &  0.73  &  0.64  &  3.09   & 1.34  &  1.54 & 1.16\\ 
    \hline

    \hline
  \end{tabular}
    
\end{table*}

\setlength{\arrayrulewidth}{.1mm}
\setlength{\tabcolsep}{5.5pt}
\renewcommand{\arraystretch}{1.3}

\begin{table*}[t]
  \centering
  \caption{Performance comparison in terms of AAE among different existing method.}
  \label{table:comparison_paper}
  \begin{tabular}{c|c|c|c|c|c|c|c|c|c|c|c|c|c}

    \hline
    Subject No. & 1 & 2 & 3 & 4 & 5 & 6 & 7 & 8 & 9 & 10 & 11 & 12 & Average\\
    
    \hline
    
    TROIKA[..]  & 2.29 & 2.19 & 2.00 & 2.15 & 2.10 & 2.76 & 1.67 & 1.93 & 1.86 & 4.70 & 1.72 & 2.84  & 2.42 (SD=2.47)\\    
    \hline
    
    JOSS[..]   &  1.33 & 1.75 & 1.47 & 1.48 & 0.69 & 1.32 & 0.71 & 0.56 & 0.49 & 3.81 & 0.78 & 1.04 & 1.28 (SD=2.61)\\
    \hline
    SPECTRAP[..] & 1.18 & 2.42 & 0.86 & 1.38 & 0.92 & 1.37 & 1.53 & 0.64 & 0.60 & 3.65 & 0.92 & 1.25 & 1.50 (SD=1.95)\\
	\hline
	
	COMB[..] & 2.06 & 3.59 & 0.92 & 1.54 & 0.97 & 1.64 & 2.25 & 0.63 & 0.62 & 4.62 & 1.30 & 1.80 & 1.82 (SD=NA)\\
    \hline
    Proposed method   &     1.16  &  1.16  &  0.79  &  0.87    & 0.71  &  1.14  &  0.71  &  0.73  &  0.64  &  3.09   & 1.34  &  1.54 & 1.16 (SD=1.74)\\
    \hline
  \end{tabular}
  
\end{table*} 

In Table \ref{table:comparison}, the AAE values obtained from 12 subjects namely (subj1,subj2,...,subj12) are presented for the four cases mentioned above. The average of AAEs for all 12 subjects is also reported in the last column of the table for each cases. The condition when AAE$>10$ indicates severe failure of the heart rate estimation, which will result in very poor heart rate tracking. It is observed from the table that for some subjects (subject-2, subject-3, subject-6), method C-1 provides AAE$>10$ $(43.57,50.08,54.68)$. However, if we consider the case C-2, such failure never occurs and in most of the cases, in comparison to C-1, the AAE values are reduced. Interestingly, the overall performance improves drastically. In this case, the lower AAE is obtained with very low standard deviation(SD). It can be easily observed that use of LMS instead of RLS can not provide better performance. In this case, both performance indices (AAE and SD) increase.  \\

In order to compare the performance of the proposed method with that obtained by some of the existing methods, in Table \ref{table:comparison_paper}, AAE values obtained for all 12 subjects are reported for JOSS, TROIKA, SPECTRAP, COMB and proposed method. It is to be noted that in case of JOSS and TROIKA methods, for performance evaluation, some initial time frames are excluded. However, in the proposed method, no time frames are excluded. It is found that the performance of the proposed method is superior in most of the subjects in comparison to that obtained by other three methods. Among 12 subjects, only in four cases (subject 8, 9, 11, 12), it is observed that the AAE values obtained by the proposed method are slightly higher but within the acceptable limit. However, the average AAE over all 12 subjects obtained by the proposed method is found lowest among all methods. In this table, in the last column along with average AAE, the average value of standard deviation over all 12 subjects are also reported, which is found to be lowest in case of the proposed method. \\

\begin{figure}[h]
  \centering
  \includegraphics[width=3.5in, height = 2.5 in]{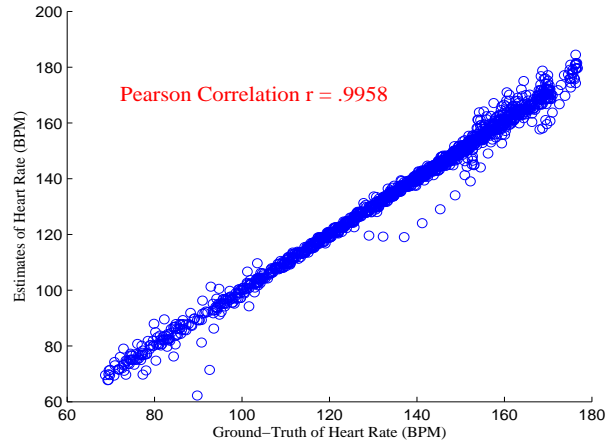}\\
  \caption{The Pearson correlation of the estimation results on the 12 dataset.}\label{Fig:pearson}
\end{figure}

\begin{figure}[h]
  \centering
  \includegraphics[width=3.5in, height = 2.5 in]{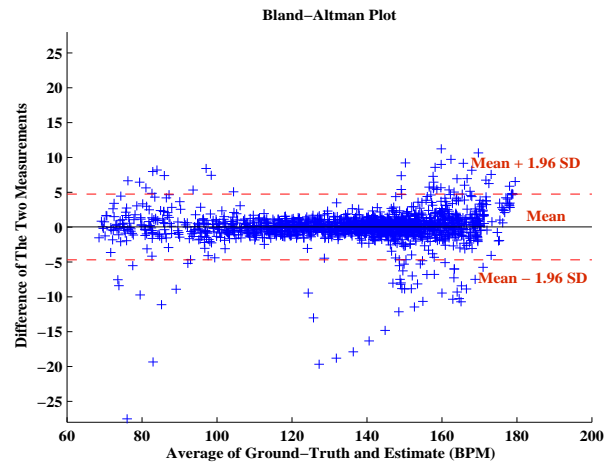}\\
  \caption{The Bland-Altman plot of the estimation on the 12 dataset.}\label{Fig:bland}
\end{figure}
 
In view of further investigating the quality of the heart rate estimation obtained by the proposed method, in Fig. \ref{Fig:pearson}, estimated heart rate is plotted against the ground truth considering all time frames of all 12 subjects. It is observed from the figure that a linear relation exists between ground truth and estimated heart rate and the approximated linear curve passes close to the origin.
The Pearson values of correlation coefficient is found 0.9958 which indicates a very consistent estimation. As it is almost nearly 1, it indicates the validation of highly accurate estimation of heart rate. Next, using all time frames of all 12  subjects Bland-Altman plot is shown in Fig. \ref{Fig:bland}. It is found that a reasonable limit of agreement (LOA) [-4.26 4.63] is obtained when 95\% data exist within 1.96$\sigma$. No matter whether the ground truth is very small or large, the difference between estimated values and ground truth is found within a satisfactory limit. It can be inferred from Table \ref{table:comparison} that, if estimation accuracy is the main concern, the proposed method can be implemented with confidence and if faster calculation is main concern, instead of RLS algorithm LMS along with the tracking step can be implemented sacrificing estimation accuracy. 

\begin{figure*}[h]
  \begin{minipage}[b]{0.48\linewidth}
 	 \centering
  	\centerline{\includegraphics[ width = 3.5in, height = 2.7 in]{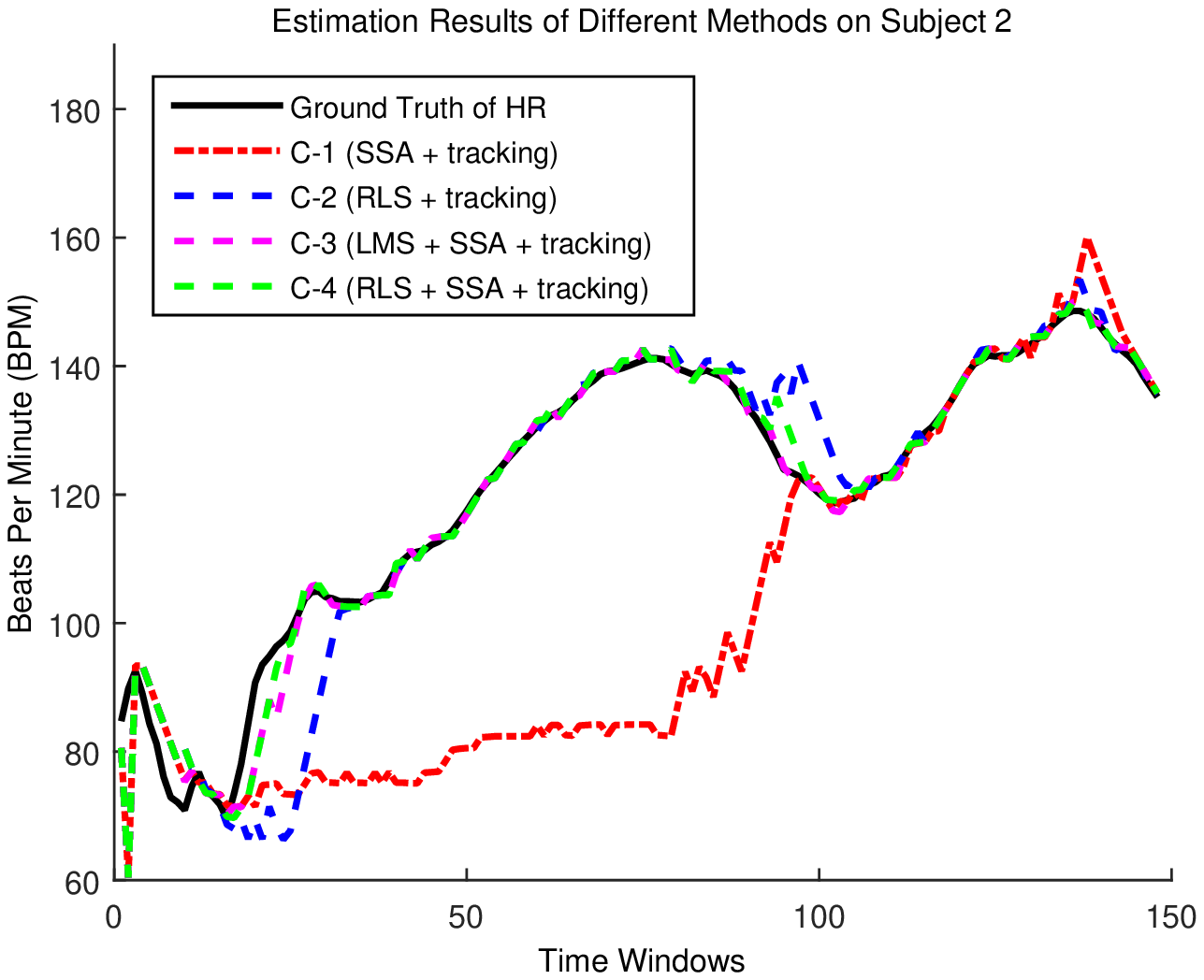}}
  	\centerline{\footnotesize{(a)}}\medskip
  \end{minipage}
  \begin{minipage}[b]{0.48\linewidth}
 	 \centering
  	\centerline{\includegraphics[ width = 3.5in, height = 2.7 in]{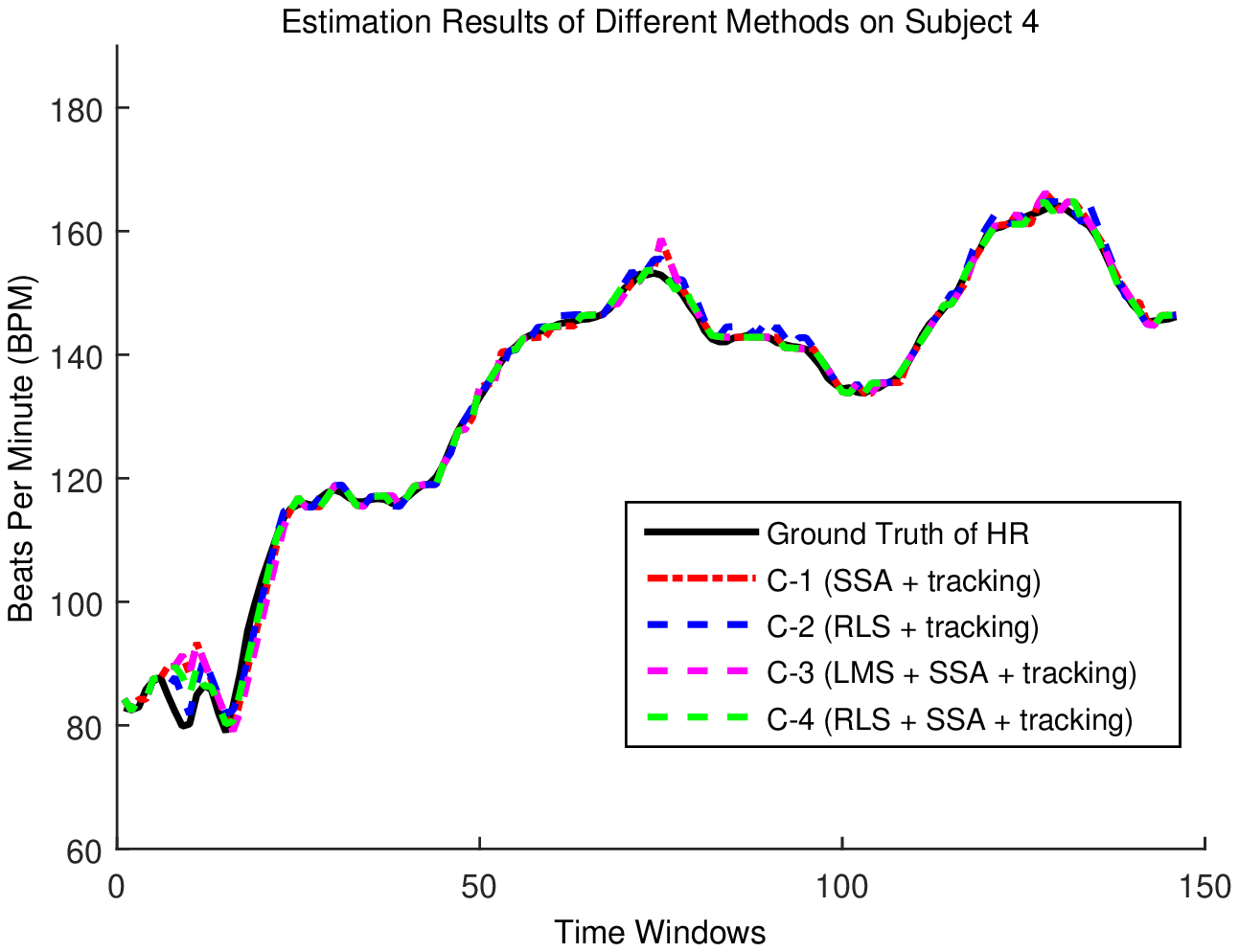}}
  	\centerline{\footnotesize{(b)}}\medskip
  \end{minipage}
  
  \caption{Performance comparison in terms of heart rate tracking on (a) subject 2 and (b) subject 4.}\label{Fig:comp}
\end{figure*}

It has been observed that, in general the lower the AAE of the estimated heart rate the smoother the heart rate tracking. In order to investigate the heart rate tracking capability of different approaches, recording of subject 2 and subject 4 are chosen. In Fig. {\ref{Fig:comp}}(a), performance comparison in terms of heart rate tracking is demonstrated for subject 2. In this figure, during different time interval, walking or running speed is varied. It is observed from the figure that the case C-1 fails to track right after few seconds and generates large errors which tracks completely in a different path. As expected, during initial 12 seconds, all methods face problem in tracking. However, after few seconds, better tracking performance is obtained. Between the cases C-2 and C-3, it is found that tracking error is lower in case of C-3. It is clearly observed that among all approaches, the proposed C-4 provides the best tracking performance. A similar observation can be made for the tracking performance for subject 4 in Fig. {\ref{Fig:comp}}(b). It can be observed from the figure that the proposed C-4 closely follows the ground truth while other approaches show some outliers.

In order to evaluate the performance of the proposed method, average computational time required for each time window is computed by using Matlab2012b on a Core i7 4770K 3.6 GHz processor equipped with 16 GB RAM with Windows 10. The average computational time for each time window required by the proposed method has been found to be 170 ms. It is to be mentioned that the codes are not fully optimized. However, the time required by the proposed method is satisfactory and within the acceptable limit for online operation.


\section{Conclusion}
In this paper, an efficient scheme for heart rate estimation during intensive physical exercise from wrist type PPG signal is developed based on motion artifact reduction in both time and frequency domain. In the proposed time domain scheme, unlike conventional adaptive filter based motion artifact reduction methods, acceleration data from three channels are separately used in three stage cascaded RLS filtering. It is found that instead of using combined three channel acceleration data as a single reference for noise cancellation in RLS filter, sequential use of X, Y and Z channel data as reference to cascaded RLS filters  can provide significant reduction of motion artifact from PPG signal. In frequency domain, the SSA based noise reduction scheme is implemented. The PPG signal obtained at the output of the proposed multistage RLS block is logically combined with the SSA stage output to obtain better motion artifact reduction. Finally, heart rate is computed from spectral peak  of the resulting PPG signal and for a better continuous heart rate estimation, a tracking algorithm is incorporated considering the neighbouring estimations. It is observed that in some cases, SSA based approach exhibits complete failure in heart rate estimation, whereas the proposed RLS based scheme offers consistent estimates with acceptable level of accuracy in all cases. However, conditional incorporation of SSA technique along with the RLS based approach provides very satisfactory performance in terms of  average absolute BPM error and standard deviation. From extensive simulation on several PPG data, it is observed that, the proposed method is capable of tracking the ground truth with high estimation accuracy regardless the abrupt change in heart rate with respect to time.

\renewcommand{\baselinestretch}{1.5}
\bibliographystyle{model2-names}

\section*{References}
\bibliography{refe}

\renewcommand{\baselinestretch}{1.5}

\clearpage\newpage

\renewcommand{\footnotesize}{\fontsize{10}{9}\selectfont}

\end{document}